\documentclass[showpacs,preprintnumbers,twocolumn]{revtex4-1}
\usepackage{color,graphicx}% Include figure files
\usepackage{amsmath}
\usepackage{amssymb}
\usepackage{subfigure}
\begin{document}
\title{Dynamical approach to weakly dissipative granular collisions}
\author{Italo'Ivo Lima Dias Pinto}
\author{Alexandre Rosas}
%\email{arosas@fisica.ufpb.br}
\affiliation{Departamento de F\'{\i}sica, CCEN, Universidade Federal da Para\'{\i}ba, Caixa Postal 5008, 58059-900, Jo\~ao Pessoa, Brazil}
\author{Katja Lindenberg}
\affiliation{Department of Chemistry and Biochemistry and BioBircuits Institute, University of California San Diego, La Jolla, California 92093-0340, USA}

\date{\today}

\begin{abstract}
Granular systems present surprisingly complicated dynamics. In particular, nonlinear interactions and energy dissipation play important roles in these dynamics. Usually, constant coefficients of restitution are introduced phenomenologically to account for energy dissipation when grains collide. The collisions are assumed to be instantaneous and to conserve momentum. Here, we improve on this phenomenology by introducing the dissipation through a viscous (velocity dependent) term in the equations of motion for two colliding grains. Using a first order approximation, we solve the equations of motion in the low viscosity regime. This approach allows us to calculate the collision time, the final velocity of each grain, and a coefficient of restitution that depends on the relative velocity of the grains. We compare our analytic results with those obtained by numerical integration of the equations of motion.
\end{abstract}

\maketitle

\section{Introduction}

The characterization of granular matter is extremely broad, and includes essentially any conglomeration of discrete macroscopic particles.  These can be as small as grains of sand and as large as asteroids, they can be in a condensed or gas-like phase. The condensed phases may exhibit characteristics of solids or fluids or gases or various combinations thereof. Granular matter is important in more industrial applications than can be listed here, and exhibits a huge variety of interesting behaviors that have provided food for thought over centuries of time.  Behaviors such as the so-called jamming transition and the formation of patterns are frequently subjects of current research, as is the propagation of energy in granular materials.  

A feature common to granular matter is the fact that energy is lost every time grains collide. Indeed, the grains have to be very hard and difficult to compress for a collision not to lead to a loss of energy. Yet it is usually the case that momentum is conserved in these inelastic collisions. The conservative limit, where only elastic collisions are involved, is famously illustrated by Newton's cradle, consisting of a row of very hard balls that just touch, each hanging on a string of the same length attached to a common support.  When the ball at one end is picked up and released so that it collides with the next ball, the energy passes down the row until the last ball flies up to the same height as the first ball before it was released (the other balls remain at rest).  The last ball then flies back, the energy is transferred across the row again, and the first ball flies up to the same height~\cite{hutzler,glendinning,sekimoto}.  This continues, although not forever because of course some small amount of energy must be lost at each collision event.

The prototypical phenomenological description of energy loss involves the \emph{coefficient of restitution} $\varepsilon$ in the equation that describes a collision between two grains,
\begin{equation}
 v_{f2} - v_{f1} = \varepsilon \left ( v_{i1} - v_{i2} \right ).
 \label{restitution}
\end{equation}
Here the $v$'s represent the velocities, the subscripts $i$ and $f$ stand for initial (before collision) and final (after collision) and the numbers label particles $1$ and $2$. At each collision this description leads to an energy loss at each collision of $ 1 - \varepsilon ^2$  of the kinetic energy of the center of mass before the collision. For a successfully built Newton's cradle, $\varepsilon$ is exceedingly small.

The coefficient of restitution is usually treated as a parameter independent of the velocities. And yet, it is broadly recognized that this can not be totally correct because it leads to problematic features in the asymptotic behavior such as the so-called inelastic collapse in a granular gas because there may be an infinite number of collisions in a finite time~\cite{inelastic_collapse,Bernu}.  Indeed, when one considers realistic interaction models, it is in fact universally the case that interactions of any two compressible grains are nonlinear.  For instance, the interactions between two spherical objects obey Hertz's law, where the repulsion is proportional to the compression to the power $3/2$ rather than the more familiar Hook's law where the repulsion is simply proportional to the compression. The consequence of this non-linearity is that the duration of a collision depends on the initial velocities of the particles before the collision. Therefore, when a dissipative collision occurs, the mechanism responsible for the dissipation of energy acts for different lengths of time depending on the initial velocities, leading to distinct energy losses, and consequently to a velocity-dependent coefficient of restitution. 

In this contribution, we present a perturbative but very broadly applicable approach  to analyze the consequences of a general viscoelastic model of dissipation on the outcome of a collision in the low dissipation regime. This study has interesting implications for the understanding of a full range of realistic collisions, which should be useful for the study of granular gases and for pulse propagation in granular chains. 

%This paper is presented as follows. In Section \ref{sec:model} we describe the model of interactions between  two particles and present the perturbative theory. Finally, in Section~\ref{sec:conclusions} we draw our conclusions.

We accomplish three objectives, all in the limit of low but non-zero dissipative forces.  One is to discuss the equations of motion that describe the collision of two grains.  The scenario is this: the two grains are initially just touching.  Grain 1 has velocity $v_1$ and grain 2 has velocity $v_2$.  How these grains came to have these velocities is not important.  If our two grains are elements in a granular gas, they may, for example, have arrived at this point due to previous collisions with other grains. Or, one might be preparing an experiment with just two grains, where one grain is given a kick of some kind at time $t=0$ to cause it to start moving with velocity $v_1$ while the other is initially at rest ($v_2=0$).  The equations of motion then determine the further evolution of the two granules.

Our second objective is to calculate the duration of a collision as a function of the initial velocities.  This makes explicit our assertion that collisions are in general not instantaneous and that their duration in fact depends on the initial velocity of the colliding granules.  As pointed out above, this in turn leads to a velocity-dependent coefficient of restitution.

Our third objective is to calculate the coefficient of restitution and the final granular velocities. In the absence of dissipative forces the coefficient of restitution is equal to unity.  We are able to explicitly calculate the lowest order corrections to this and thus to obtain an explicit form for the dependence of the coefficient of restitution on the initial velocity difference of the granules.

\section{The model}
\label{sec:model}

Viscoelastic forces for the collision of two spheres include two terms. The first is due to the elastic repulsion between the two particles and has it origins in Hertz's theory~\cite{hertz,landau}. The second term stands for the viscous dissipation via a dashpot~\cite{ji,alizadeh}. Hence, the contact force can be written as
% NEW - we work on the general case, not only for spheres, so corrections
% are made in the equation and the text bellow
\begin{equation}
  F = - r \left ( x_1 - x_2 \right )^{n-1} - \gamma \left ( x_1 - x_2 \right )^\alpha \left ( \dot{x}_1 - \dot{x}_2 \right ).
 \label{viscoforce}
\end{equation}
Here $x$ is the displacement of a particle from its initial position at the beginning of a collision.  A dot denotes a derivative with respect to time, and the subscripts on $x$ label the two particles.  The coefficient
$r$ is a constant dependent on Young's modulus and Poisson's ratio, $ \gamma $ is the coefficient of viscosity, $ \alpha $ is a constant that defines the specific viscoelastic model, and $n$ depends on the topology of the contact between the particles.  It is equal to $5/2$ for spheres, but we leave it as $n$ for the sake of generality. We say that particle $1$ is to the left of particle $2$, and that our system of coordinates increases from left to right. % END of changes

\subsection{Equations of Motion}

The equations of motion for two particles of mass $m_1$ and $m_2$ during a collision are
\begin{eqnarray}
\label{eq:motion}
m_1 \ddot{x}_1 & = & - r \left( x_1 - x_2 \right)^{n-1} - \gamma \left( x_1 - x_2 \right)^\alpha \left( \dot{x}_1 - \dot{x}_2 \right), \nonumber \\
m_2 \ddot{x}_2 & = & r \left( x_1 - x_2 \right)^{n-1} + \gamma \left( x_1 - x_2 \right)^\alpha \left( \dot{x}_1 - \dot{x}_2 \right).
\end{eqnarray}
From Eq. (\ref{eq:motion}), conservation of momentum immediately follows,
\begin{equation}
\label{eq:moment_cons}
% NEW - equation corrected
m_1 \ddot{x}_1 + m_2 \ddot{x}_2  =  0,
% END of NEW
\end{equation}
so that $m_1\dot{x}_1+m_2\dot{x}_2=const$. Equation~(\ref{eq:motion}) also leads to an uncoupled equation for the difference variable $z = x_1 - x_2$,
\begin{equation}
\label{eq:dynamics}
\ddot{z}  =  - \frac{r}{\mu} {z}^{n-1} - \frac{\gamma}{\mu} z^{ \alpha }\dot{z},
\end{equation}
where $\mu$ is the reduced mass $\mu^{-1} = m_1^{-1} + m_2^{-1}$. 
For the latter equation, the initial conditions are ${z}(0) = 0$ because we deal with configurations where the granules are initially just touching each other, and ${\dot{z}}(0) = v_{i1} - v_{i2} \equiv v_0$. Here $v_{i1}$ and $v_{i2}$ are the initial velocities of the two colliding granules. 

An analytic solution ${z}(t)$ of Eq.~(\ref{eq:dynamics})
seems not to be available. However, we have been able to obtain the velocities at the end of the collision as a function of the initial velocities in the low viscosity limit. We rewrite Eq. (\ref{eq:dynamics}) as a first order differential equation of the velocity as a function of the position.  Defining $v = \dot{z}$, and noting that 
\begin{equation}
\ddot{z} = \dfrac{dv}{dt}=\dfrac{dv}{dz} \dfrac{dz}{dt} = v \dfrac{dv}{dz},
\end{equation}
we rewrite Eq. (\ref{eq:dynamics}) as
\begin{equation}
\label{eq:dvdz}
v \dfrac{dv}{dz} = - \frac{r}{\mu} z^{n-1} - \frac{\gamma}{\mu} z^\alpha v.
\end{equation}
In the absence of dissipation ($ \gamma = 0$), this equation admits two solutions for $v(z)$,
\begin{equation}
\label{eq:solutions}
v_{\pm}(z) = \pm \sqrt{{v_0}^2 - \dfrac{2 r z^n}{\mu n}}.
\end{equation}
Obviously, the positive sign should be considered during compression, and the negative one during decompression.

Our approximation in the low dissipation regime starts by writing the velocity  as a perturbation on the non-dissipative solution. Consequently, during compression we have
\begin{equation}
\label{eq:try_solution}
v(z) = v_{+}(z) + \gamma v_{C}(z),
\end{equation}
where $v_C(x)$ is a function to be determined.
Substituting the trial solution Eq. (\ref{eq:try_solution}) in Eq. (\ref{eq:dvdz}), and collecting the terms of order $\gamma$, we have
\begin{equation}
\label{eq:correction_vel}
\frac{r}{\mu} z^{n-1} v_{C} + \left( \frac{2 r z^n}{\mu n} - {v_0}^2 \right) \left( \frac{z^{ \alpha }}{\mu} + v_{C}' \right) = 0,
\end{equation}
where a prime denotes a derivative with respect to $z$. Furthermore, the condition $v_C(0) = 0$ is necessary to satisfy the initial conditions. The solution of Eq.~(\ref{eq:correction_vel}) with the initial conditions $v_C(0)=0$, when added to $v_+(z)$, then gives us the compression velocity to first order in the dissipation,
\begin{widetext}
\begin{equation}
\label{v(x)}
v(z) = \sqrt{{v_0}^2 - \dfrac{2 r z^n}{\mu n}} - \gamma \dfrac{z^{1+\alpha}\left[2(1 + \alpha ) + n\, {}_2F_1(1,\frac{1}{2}+\frac{1+ \alpha }{n};1+\frac{1+ \alpha }{n};\frac{2 r z^n}{n \mu {v_0}^2})\right]}{(1 + \alpha ) (2 + n + 2 \alpha ) \mu},
\end{equation}
\end{widetext}
where ${}_2F_1$ is a hypergeometric function~\cite{math.methods}.

\subsection{Collision Time}

Our next objective is to calculate the collision time, which we do in two parts.  First we calculate the compression time of the collision as the two grains compress one another and then, as the grains move apart, the attendant decompression time.  The collision time is then the sum of the two.

In order to calculate the compression time we must calculate the maximum compression $z_{max}$. Once again, we look for a first order correction in $\gamma$,
\begin{equation}
\label{x_max}
z_{max} = \left( \dfrac{n \mu {v_0}^2}{2 r} \right)^{1/n} \left( 1 - \gamma z_{C} \right),
\end{equation}
where $z_C$ is a constant to be determined and $\left( n \mu {v_0}^2/2 r \right)^{1/n}$ is the maximum compression in the absence of dissipation, obtained by setting $v_\pm(z)=0)$ in Eq.~(\ref{eq:solutions}). 

The relative velocity must vanish for the maximum compression. This is the point at which the colliding grains stop moving so as to increase compression and begin to separate. Therefore, we evaluate Eq. (\ref{v(x)}) at $z=z_{max}$ as given in Eq. (\ref{x_max}),  expand to first order in $ \gamma $, and set the left hand side equal to zero. 
Solving  
the resulting equation for $z_{C}$ and inserting the result in Eq.~(\ref{x_max}) we find for the maximum compression
\begin{widetext}
\begin{equation}
z_{max} =  \left( \dfrac{n \mu {v_0}^2}{2 r} \right)^{\frac{1}{n}} \left[ 1 - \gamma \dfrac{2^{-\frac{1+n+\alpha}{n}} \sqrt{\pi} \left( \frac{n \mu}{r}  \right)^{\frac{1+\alpha}{n}} {v_0}^{\frac{2-n+2 \alpha}{n}} \Gamma \left( \frac{1+\alpha}{n} \right)}{n^2 \mu \Gamma \left( \frac{3}{2} + \frac{1+\alpha}{n} \right)} \right].
\end{equation}
\end{widetext}

Next we can calculate the compression time as 
\begin{equation}
\label{integral}
T_{compression} = \int_0^{z_{max}} \! \dfrac{1}{v(z)} dz
\end{equation}
where, to order $\gamma$, $\frac{1}{v(z)}$ is obtained from Eq. (\ref{v(x)}) as
\begin{widetext}
\begin{equation}
\label{1/v(x)}
\dfrac{1}{v(z)} = 
\dfrac
{1}{\sqrt{{v_0}^2 - \frac{2 r z^n}{n \mu}}} + \gamma \dfrac{z^{1+\alpha} \left[2(1+\alpha) + n\, {}_2F_1(1,\frac{1}{2}+\frac{1+\alpha}{n};1+\frac{1+\alpha}{n};\frac{2 r z^n}{n \mu {v_0}^2})\right]}
{ (1+\alpha)\left( 2 +n+2\alpha\right)\mu(v_0^2 - \frac{2 r z^n}{n\mu} )}.
\end{equation}
\end{widetext}
Substituting this into Eq.~(\ref{integral}) and integrating leads to a contribution of order $\gamma^0$, and a cancellation of two terms of order $\gamma^{1/2}$.
Consequently, the compression time does not show any $ \gamma $ dependence up to first order, that is, to first order it is independent of the viscosity:
%\begin{widetext}
\begin{eqnarray}
%\begin{equation}
\label{time_comp}
T_{compression} &=& \dfrac{2^{-\frac{1}{n}} n^{-1+\frac{1}{n}} \sqrt{\pi} r^{- \frac{1}{n}} {\mu}^{\frac{1}{n}} {v_0}^{-1 + \frac{2}{n}} \Gamma(\frac{1}{n})}{\Gamma(\frac{1}{2} + \frac{1}{n})}\nonumber\\ ~~~~&& + \mathcal{O}({\gamma}^{3/2}).
\end{eqnarray}
%\end{widetext}

Next we move on to the decompression which, as we noted above, starts with $v = 0$ and $z = z_{max}$ (maximum compression). It ends when $z = 0$ (grains lose contact). Following the same steps used to determine the compression time, we assume a perturbative solution
\begin{equation}
\label{v_format}
v(z) = v_{-}(z) + \gamma v_{u}(z),
\end{equation}
substitute it into Eq. (\ref{eq:dvdz}), and expand the latter up to first order in $\gamma$. We find
\begin{eqnarray}
\label{eqvcorrect2b}
v(z) = \dfrac{C}{\sqrt{n \mu {v_0}^2 - 2 r z^n}} - \dfrac{ 4 r z^{1+n+\alpha}}{\left( 2 + n + 2 \alpha \right) \mu \left( 2 r z^n - n \mu {v_0}^2 \right)} \nonumber\\
- \dfrac{z^{1+\alpha} \left( 2 + 2 \alpha + n {}_2F_1 \left( 1,\frac{1}{2} + \frac{1 + \alpha}{n};1+\frac{1 + \alpha}{n};\frac{2 r z^n}{n \mu {v_0}^2} \right) \right)}{\left( 1 + \alpha \right) \left( 2 + n + 2 \alpha \right) \mu}.\nonumber\\
\end{eqnarray}
Here $C$ is a constant to be determined by the continuity of the solutions (\ref{eqvcorrect2b}) and (\ref{v(x)}) at $z_{max}$. Remembering that $v(z_{max}) = 0$, where $z_{max}$ is given by Eq. (\ref{x_max}), and expanding up to first order in $\gamma$, $C$ is found to be
\begin{eqnarray}
\label{C_1}
C &=& 
\frac{2^{-\frac{\alpha +1}{n}} \sqrt{\pi } (n \mu)^{\left(-\frac{1}{2}+\frac{\alpha +1}{n}\right)}  r^{-\frac{\alpha +1}{n}} v_0^{\frac{n+2 \alpha +2}{n}} \Gamma \left(\frac{\alpha +1}{n}\right)}{\Gamma \left(\frac{\alpha +1}{n}+\frac{3}{2}\right)}.\nonumber\\
\end{eqnarray}
Substituting Eq. (\ref{C_1}) into Eq. (\ref{eqvcorrect2b}) for $z=0$ (end of collision) and expanding in series up to first order in $\gamma$, we obtain
\begin{equation}
\label{v_final}
v_{final} = -v_0 + \gamma \dfrac{ \left( \frac{n \mu v_0^2}{2r} \right)^{\frac{1+\alpha}{n}} \sqrt{\pi} \Gamma (1 + \frac{1+\alpha}{n})}{\mu \Gamma (\frac{3}{2} + \frac{1+\alpha}{n})}.
\end{equation}
Furthermore, since the zero-th order term of the decompression velocity Eq. (\ref{v_final}) is the negative of the compression velocity Eq. (\ref{v(x)}), and the limits of integration of the compression and decompression times are switched, the decompression time is the same as the compression time up to terms of order ${\gamma}^{3/2}$. Hence the total collision time is twice $T_{compression}$,
\begin{eqnarray}
\label{time_total}
T = \dfrac{2^{1-\frac{1}{n}} n^{-1+\frac{1}{n}} \sqrt{\pi} r^{- \frac{1}{n}} {\mu}^{\frac{1}{n}} {v_0}^{-1 + \frac{2}{n}} \Gamma(\frac{1}{n})}{\Gamma(\frac{1}{2} + \frac{1}{n})} + \mathcal{O}({\gamma}^{3/2}),\nonumber\\
\end{eqnarray}
and we have arrived at our second objective.

\begin{figure}
  \includegraphics[width=9cm]{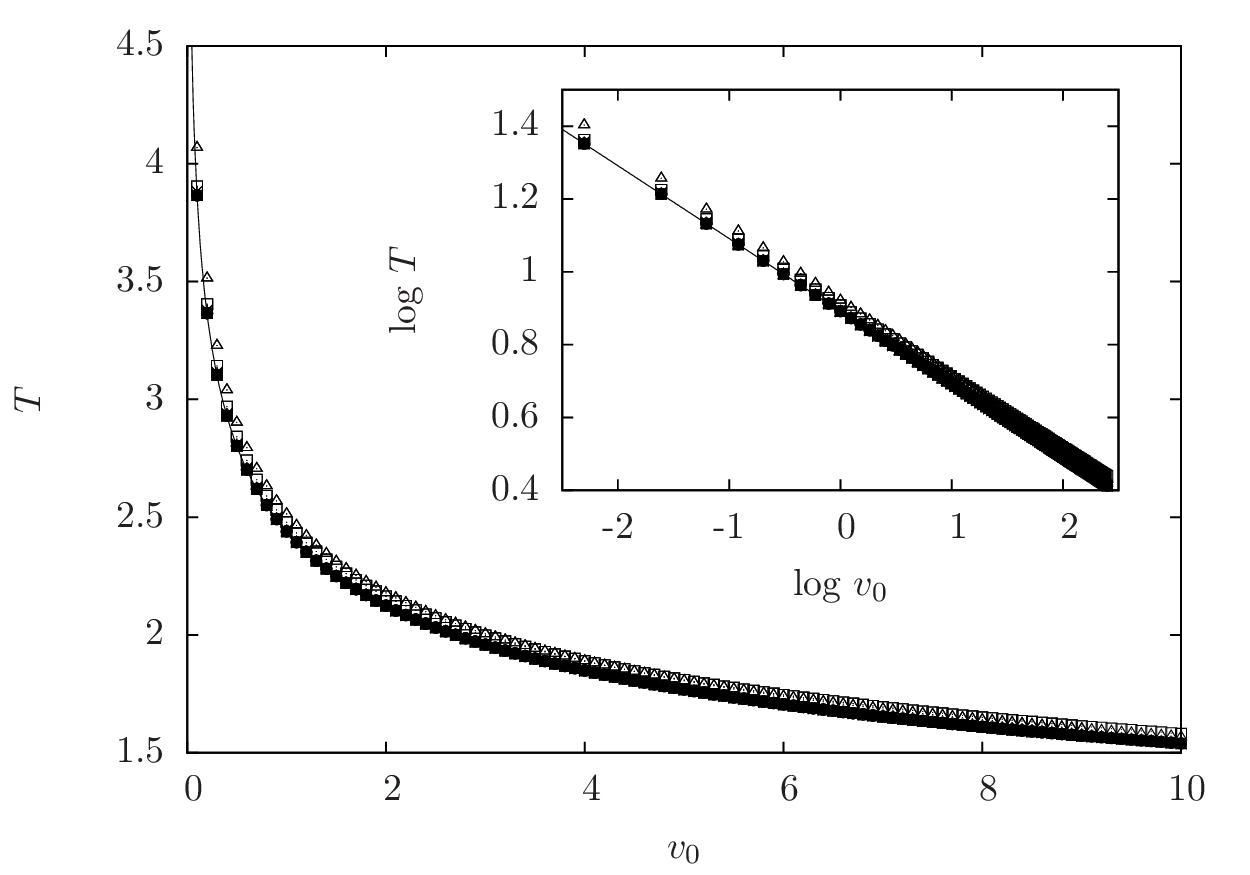}
  \caption{Collision duration for several values of $ \gamma $ (0.0001, 0.001, 0.01 and 0.1) and two different values of $ \alpha $ (0 and $1/2$), obtained via numerical integration of the equations of motion. The line represent the theoretical prediction of Eq. (\ref{time_total}). The inset shows the same data on a log-log scale. In this figure, the parameters are as follows: $m_1 = m_2 =1, \; R_1 = R_2 = 1$. 
 As predicted, for these values of $\gamma$ and 
 $\alpha$ the collision durations are essentially independent of these parameters. 
 \label{fig:time}}
\end{figure}

In a recent paper, the merits and problems of different choices of the parameter $ \alpha $, and even a generalization of the above model, were discussed~\cite{alizadeh}. Here, we consider two choices of this parameter that have been commonly used in the literature to test Eq.~(\ref{viscoforce}). The simplest case, $ \alpha =0$, was proposed in~\cite{lee}, while the case $ \alpha = 1/2$, proposed independently in~\cite{kuwabara,brilliantov}, seems to be more widespread in the granular gas community. In Fig.~\ref{fig:time} we show the duration of the collision for two equal spherical grains ($n=5/2$) as a function of the relative velocity at the beginning of the collision for both values of $ \alpha $. As can be seen in the figure, for small $ \gamma $ the data is very well predicted by our approximation, independently of the exponent $ \alpha$ (the approximation only starts to deviate from the theoretical prediction for $ \gamma = 0.1$, represented by the filled circles). This $ \alpha-$independence of the duration of the collision is one of the striking predictions of our theory. Another interesting characteristic of our solution is the power-law dependence of $T$ on the initial relative velocity $v_0$, as evidenced in the inset of Fig.~\ref{fig:time}.

\begin{figure}
  \includegraphics[width=9cm]{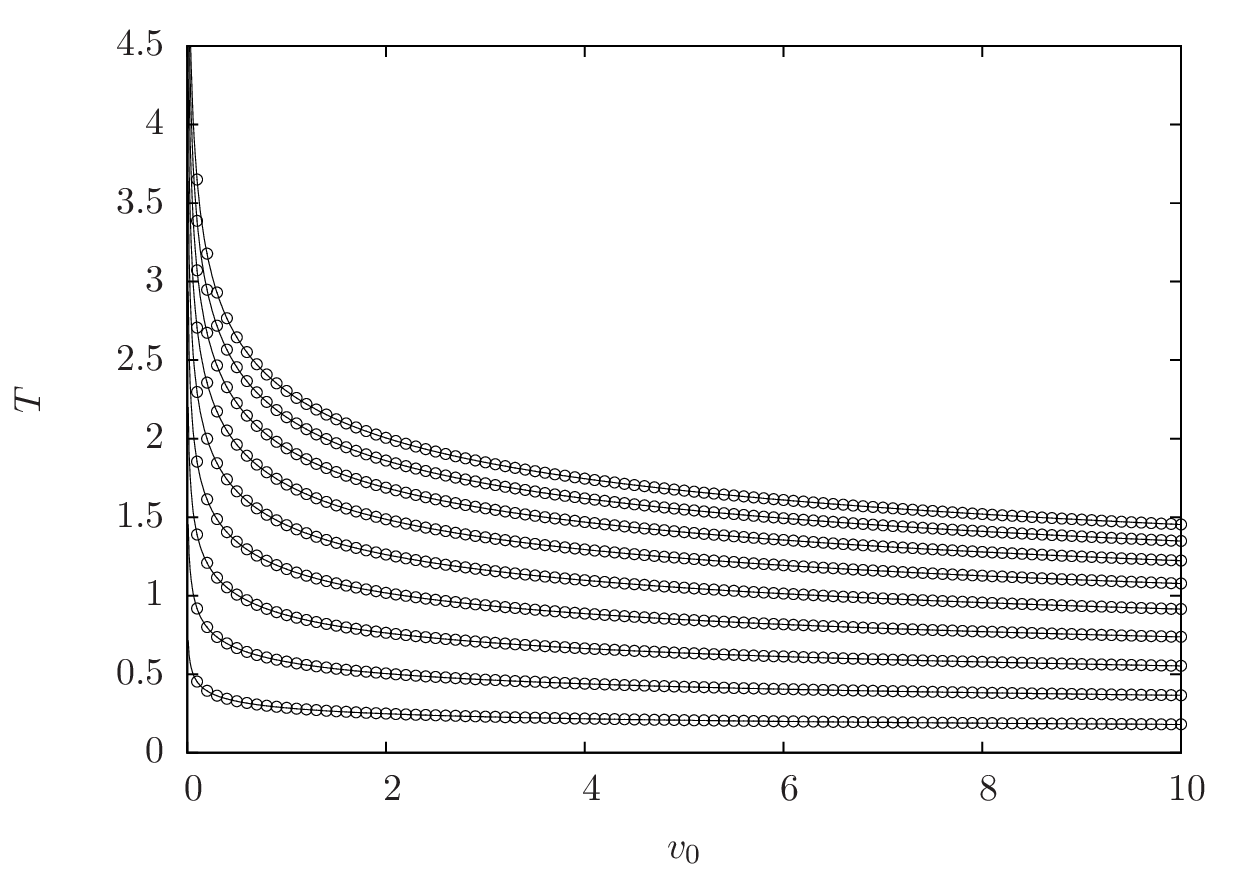}
  \caption{Duration of collisions for several values of $ r_2 $ (from bottom to top, $ r_2 $ varies from 0.1 to 0.9 in steps of 0.1) obtained via numerical integration of the equations of motion. The lines represent the theoretical prediction Eq.~(\ref{time_total}). In this figure the parameters are as follows: $m_1 = 1, \; R_1  = 1, \; \alpha = 0.5$ and $ \gamma = 0.001.$\label{fig:timeR}}
\end{figure}

In Fig.~\ref{fig:timeR} we again show $T$ as a function of $v_0$, but this time for particles of different sizes. Fixing the radius of granule 1 and varying the radius of granule 2 (assuming that they have the same density), we can see that the collision takes longer for larger values of $r_2$.
% NEW
\subsection{Final Velocity and Coefficient of Restitution}
We next turn our attention to the final velocities. From the conservation of momentum, we know that the total momentum at the beginning and end of the collision must be the same, 
\begin{equation}
  m_1 v_{i1} + m_2 v_{i2} = m_1 v_{f1} + m_2 v_{f2},
\end{equation}
where the subscripts $i$ and $f$ once again label the initial and final velocities of the grains. On the other hand, we also know from the definition of $z(t)$ that
\begin{equation}
  \dot{z}(T) = v_{final} = v_{f1} - v_{f2}.
\end{equation}
Solving the set of the two equations above for $v_{f1}$ and $v_{f2}$ we find
\begin{eqnarray}
\label{finalvelocities}
  v_{f1} &=& \frac{m_1 - m_2}{m_1 + m_2} v_1 + \frac{2 m_2}{m_1 + m_2} v_2 \nonumber \\
            +&& \frac{2^{-\frac{\alpha +1}{n}} m_2}{m_1 + m_2} \frac{\sqrt{\pi } \Gamma \left(\frac{\alpha +1}{n}\right) \left(\frac{r}{\mu  n}\right)^{1-\frac{\alpha +1}{n}}}{r \Gamma \left(\frac{\alpha +1}{n}+\frac{3}{2}\right)} \left(v_{i1}-v_{i2}\right)^{\frac{2 (\alpha +1)}{n}} \gamma,\nonumber \\ \\
  v_{f2} &=& - \frac{m_1 - m_2}{m_1 + m_2} v_2 + \frac{2 m_1}{m_1 + m_2} v_1 \nonumber \\
            -&& \frac{2^{-\frac{\alpha +1}{n}} m_1}{m_1 + m_2} \frac{\sqrt{\pi } \Gamma \left(\frac{\alpha +1}{n}\right) \left(\frac{r}{\mu  n}\right)^{1-\frac{\alpha +1}{n}}}{r \Gamma \left(\frac{\alpha +1}{n}+\frac{3}{2}\right)} \left(v_{i1}-v_{i2}\right)^{\frac{2 (\alpha +1)}{n}} \gamma .\nonumber\\ 
% \label{finalvelocities)
\end{eqnarray}
As expected, the result for an elastic collision is recovered when $\gamma=0$. Further, the influence of the dissipation is greater on the lighter particle, and the influence of the initial condition on the change in the final velocities due to dissipation depends only on the relative velocity.

\begin{figure}
  \includegraphics[width=9cm]{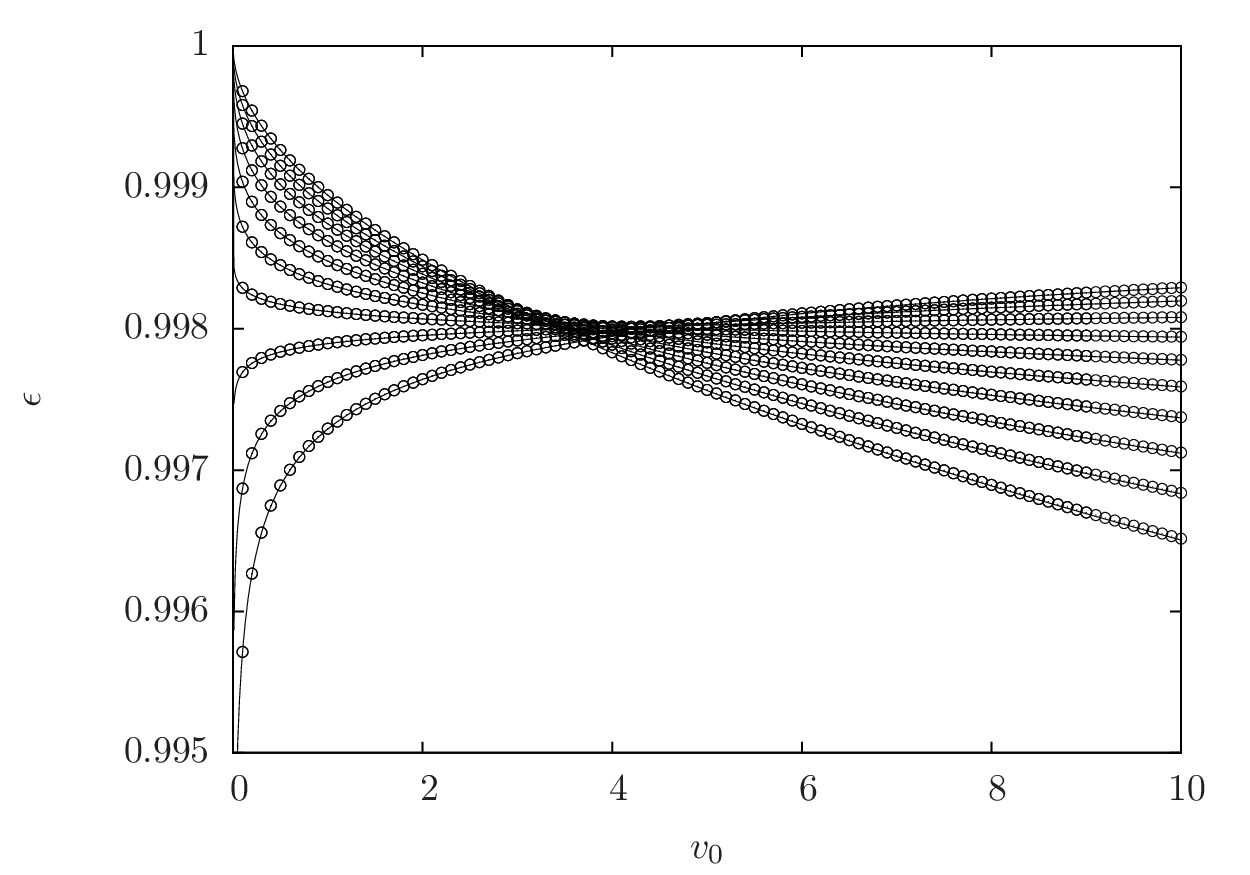}
  \caption{Coefficient of restitution as a function of the initial relative velocity. Each curve corresponds to a different value of $ \alpha $ (from bottom to top on the left side, $ \alpha $ varies from 0 to 0.9 in steps of 0.1). The other parameters are: $ \gamma = 0.001, \; m_1 = m_2 = 1$ and $r_1 = r_2 = 1.$ The lines correspond to the theoretical prediction of Eq. (\ref{eq:restitution}).\label{fig:restitution}}
\end{figure}

We conclude this section by using the above results in Eq.~(\ref{restitution}) to calculate the coefficient of restitution and thus completing our third and principal objective:
\begin{equation}
  \label{eq:restitution}
  \varepsilon = 1 -  \gamma\frac{2^{-\frac{\alpha +1}{n}} \sqrt{\pi } \Gamma \left(\frac{\alpha +1}{n}\right) \left(\frac{r}{\mu  n}\right)^{1-\frac{\alpha +1}{n}}}{r \Gamma \left(\frac{\alpha +1}{n}+\frac{3}{2}\right)} \left(v_{i1}-v_{i2}\right)^{\frac{-n+2 \alpha +2}{n}}.
\end{equation}
In Fig.~\ref{fig:restitution} we show the coefficient of restitution for several values of $ \alpha. $  The agreement is equally good for all of them. An important characteristic of $ \varepsilon $ is that the its qualitative dependence on the initial relative velocity is drastically different for $ \alpha $ larger than or smaller than $ (n - 2)/2$ (in the case of spheres, this values is $1/4$).  $\alpha$ smaller than this value leads to the unphysical situation of negative coefficients of restitution for very small relative velocities. For larger $ \alpha $, the collision approaches the elastic case for small relative velocities.

\begin{figure}
\subfigure[fig:restRalpha0][$ \alpha = 0$]{\includegraphics[width=9cm]{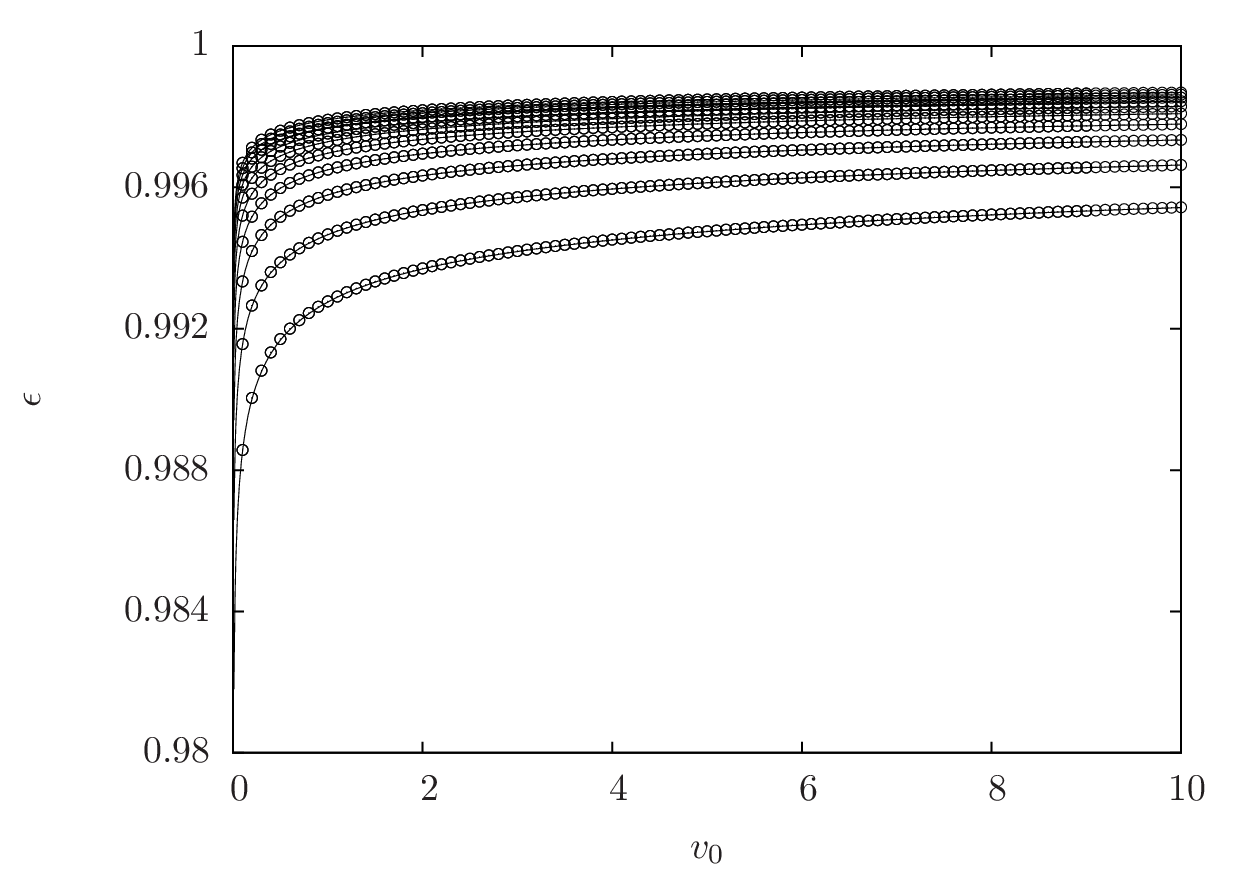}} \\
\subfigure[fig:restRalphahalf][$ \alpha = 1/2$]{\includegraphics[width=9cm]{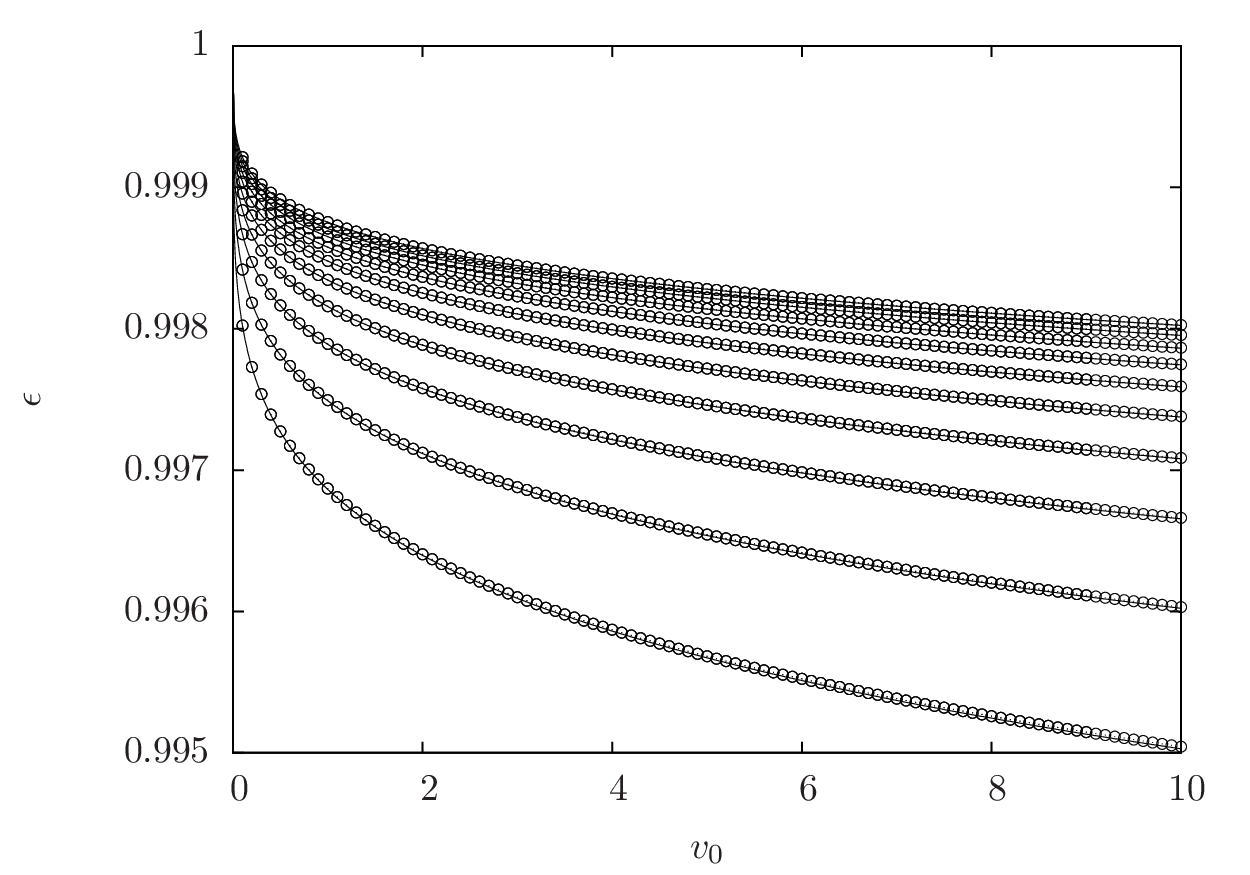}} 
 \caption{Coefficient of restitution as a function of the initial relative velocity. Each curve corresponds to a different value of the radius of granule 2 (from bottom to top, $ r_2 $ varies from 0.5 to 1.4 in steps of 0.1). The other parameters are: $ \gamma = 0.001, \; m_1 = 1$ and $r_1 = 1.$ The lines correspond to the theoretical prediction of Eq. (\ref{eq:restitution}).\label{fig:restR}}
\end{figure}

In Fig.~\ref{fig:restR}, we show the coefficient of restitution as a function of $v_0$ for particles of different sizes (same density) for $ \alpha = 0$ and $ \alpha = 1/2.$ It is evident from the figure that $ \varepsilon$ increases with the radius in both cases. However, the qualitative behavior is independent of the sizes of the grains.

\section{Conclusions}
\label{sec:conclusions}

We have succeeded in calculating quantities that characterize the collision of two granules that lead to the loss of energy (but not momentum) to the environment via viscous dissipation.  We started with an equation of motion (Newton's Law) containing a kinetic energy contribution, a force due to the elastic repulsion between the two granules, and a dashpot viscous dissipation term.  In addition to parameters related to the shape and size of the granules, the model contains two important parameters: a coefficient of viscosity $\gamma$, and a constant $\alpha$ that defines the specific viscoelastic model, cf. Eq~(\ref{viscoforce}). Our calculations are perturbative in the coefficient of viscosity, that is, we present lowest order corrections to elastic (energy-conserving) collisions.

A collision begins with the two granules just touching head-on toward each other with a relative velocity $v_0$. This velocity and configuration define the collision strength.  The collision begins at this initial moment with compression of the granules  until their relative velocity is zero (at which point the compression is a maximum). Decompression then follows, until the granules just stop touching, at which point the collision ends.  

Integration of the equations of motion leads to analytic results for several important quantities usually specified simply as phenomenological parameters.  The first is the relative velocity of the granules during compression and during decompression. We calculate the final relative velocity as a function of the separation of the centers of the granules and find the dependence on initial relative velocity and on the parameters $\gamma$ and $\alpha$, cf. Eq.~(\ref{v_final}); if the collision were elastic, the final and initial relative velocities would of course just be the negatives of one another. A second set of useful results are the final velocities of each grain, for which we obtain explicit expressions as a function of the parameters and of the initial velocities of each grain, cf. Eq.~(\ref{finalvelocities}). These are important for simulations of granular gases.

The third quantity we calculate is the duration of the collision, cf. Eq.~(\ref{time_total}). In many phenomenologies, collisions are assumed to be instantaneous. Collision are of course not instantaneous. In addition and unexpectedly, we find that while the compression and decompression times as well as their sum do depend on the initial relative velocity of the grains, to lowest order in $\gamma$ these times are independent of $\gamma$ and of $\alpha$. 

Finally, we find an analytic expression for the most important quantity in these calculations, namely, the coefficient of restitution $\varepsilon$ defined in Eq.~(\ref{restitution}). This coefficient, usually chosen phenomenologically, recognizes the inelasticity of granular collisions. We have found the dependence of this coefficient on particle shape (via the exponent in the force that determines the topology of the contact between the granules), the coefficient dependent on Young's modulus and Poisson's ratio, and, most importantly, on initial relative velocity and on the parameters $\gamma$ (to lowest order) and $\alpha$ that define the viscoelastic model, cf. Eq.~(\ref{eq:restitution}). If $\gamma=0$ the coefficient of restitution is unity, that is, there is no energy loss in the collision.  Similarly, if the initial velocities of the two granules are equal, the coefficient is trivially unity again. The dependences on these quantities are non-trivial and, we submit, essentially impossible to arrive at phenomenologically.  This then provides a physical basis for the usual phenomenological choice $\varepsilon<1$.

In this paper we have only dealt with two colliding granules, taking into account the energy loss due to an explicit viscoelastic force in the equations of motion.  This renders our results immediately applicable to granular gases where at low densities binary collisions are the most common interactions. The generalization to a granular chain or to even higher dimensional granular arrays is not trivial, but is now made considerably easier by the fact that we have explicitly found the principal ingredient of the problem.  There is nevertheless a great deal of work to be done toward these generalizations. 

\subsection*{Aknowledgements}
I. L. D. Pinto and A. Rosas aknowledge the CNPq and Bionanotec-CAPES for financial support.

\end{document}